\documentclass{PoS}
\usepackage{multirow}
\usepackage{array}
\usepackage{boldline}
\newcolumntype{?}{!{\vrule width 2pt}}
\title{Overview of Gluon Helicity Measurements at STAR}

\ShortTitle{Overview of Gluon Helicity Measurements at STAR}

\author{\speaker{Christopher Dilks}\\
        for the STAR Collaboration\\
        Pennsylvania State University\\
        E-mail: \email{cjd5150@psu.edu}}

\abstract{ 
The contribution to the spin of the proton from the gluon helicity is starting to come into
focus: for gluons carrying a large fraction $x$ of the proton momentum, evidence of positive gluon
polarization has been observed, via measurements of the longitudinal double-spin asymmetry $A_{LL}$ for
inclusive jet and dijet production. $A_{LL}$ is sensitive to the polarized gluon distribution function,
$\Delta g(x)$, and while it is positive at high $x$, it is not well constrained for $x<0.05$. Recent
measurements at STAR of observables originating dominantly from quark-gluon and gluon-gluon subprocesses
aim to improve the precision of $\Delta g(x)$ at high $x$, as well as for the first time provide insight
into the low-$x$ contribution.  $A_{LL}$ measurements of inclusive jets and dijets at midrapidity
$(|\eta|<1)$ and intermediate rapidity $(0.8<\eta<2)$ at STAR at $\sqrt{s}=200$ and $510$ GeV will be
shown, along with the statuses of ongoing analyses; these measurements will help improve the $\Delta
g(x)$ precision for $x\gtrsim 0.01$.  Recent $\pi^0$ $A_{LL}$ measurements in the forward region
$(2.65<\eta<3.9)$ at $\sqrt{s}=510$ GeV will also be presented, which probe $\Delta g(x)$ down to
$x{\sim}10^{-3}$.  Comparisons of these results to recent global analyses and extrapolations will be discussed.
}

\FullConference{23rd International Spin Physics Symposium - SPIN2018 -\\
		10-14 September, 2018\\
		Ferrara, Italy}

\newcommand{\asym}{A_{LL}}
\newcommand{\helicity}{\Delta g(x)}

\begin{document}

\section{Introduction}
The composition of the spin of the proton from the spin and orbital angular momenta of its quarks and
gluons is starting to come into focus. By the Jaffe-Manohar spin sum rule \cite{JMref}, the spin $S_p$ of
the proton is expressed as 
\begin{equation}
S_p=\frac{1}{2}=\frac{1}{2}\Delta\Sigma+\Delta G+L,
\label{JMrule}
\end{equation}
where $\Delta\Sigma$ is the fractional contribution from the quark spin, $\Delta G$ is that from the
gluon helicity, and $L$ is that from the partonic orbital angular momentum. 
The contributions $\Delta\Sigma$ and $\Delta G$ are integrals over $x$ of the respective helicity
distributions, where $x$ is the fraction of the proton momentum carried by the parton.
The gluon helicity distribution $\helicity$ is the difference between the gluon parton
distribution function (PDF) for the case of the gluon and proton helicities aligned, minus that of the
helicities anti-aligned, and the (anti)quark helicity distributions are also similarly defined.

The quark contribution is
constrained to $\Delta\Sigma\sim0.24$ \cite{Sigma1,dssv08,Sigma3,Sigma4,Sigma5}. For gluons carrying a
large fraction of the proton momentum, $x>0.05$, the gluon helicity contribution is $\Delta
G=0.23\pm0.06$ \cite{nnpdf1} and $\Delta G=0.20^{+0.06}_{-0.07}$ \cite{dssv14}; however, $\Delta G$ from
gluons with $x<0.05$ is still highly unconstrained and the full gluon helicity integrated over all $x$ is
not yet well understood \cite{nnpdf1,dssv14}. Moreover, since gluons are vastly dominant in the
$x<0.05$ region, it is important to understand their possible contribution to the proton spin. Finally,
the partonic orbital angular momentum contribution $L$ has not yet been measured.

The gluon helicity is accessible via measurements of the longitudinal double-spin
asymmetry $A_{LL}$, defined as
\begin{equation}
\asym=\frac{\sigma_{++}-\sigma_{+-}}{\sigma_{++}+\sigma_{+-}},
\label{allDef}
\end{equation}
where $\sigma_{++}$ ($\sigma_{+-}$) denotes a production cross section, given the colliding proton
helicities were the same (opposite). At STAR, $A_{LL}$ is measured for inclusive jet, dijet, and pion
production.  Recent STAR measurements of $A_{LL}$ in longitudinally polarized proton-proton
scattering aim not only to tighten constraints on the magnitude and shape of $\helicity$ for $x>0.05$,
but also aim to probe down to the largely unconstrained low-$x$ region, down to $x\sim10^{-3}$
\cite{nextGlobalAnalysis,forwardPions,intermediateDijets}.

The proton-proton scattering cross section is collinearly factorizable (see, {\it e.g.},
\cite{prospects}), with the initial state protons having contributions from the PDFs $f(x)$ for each
parton type $f$ \cite{refPDF}; if the protons are longitudinally polarized, helicity distributions
$\Delta f(x)$ contribute to the spin-dependent cross sections \cite{dssv08,nnpdf1,dssv14}. The hard
parton scattering process $f_1f_2\to f_3f_4$ is modelled by the hard-scattering cross section
$\hat{\sigma}(k_i)$ and the parton-level longitudinal double-spin asymmetry $\hat{a}_{LL}(k_i)$, which
are both calculable in perturbative QCD and are dependent on the parton momenta $k_i$
\cite{babcock,craigie}. Finally, for the production of a hadron $h$ from a scattered parton $f_3$, there
is an additional contribution from the fragmentation function $D^h_{f_3}(z)$, where $z$ is the fraction
of the momentum of $f_3$ carried by $h$ \cite{kaoFrag}. 

In terms of
these distributions, $\asym$ for $h$ production is expressed as 
\begin{equation}
\asym=
\frac{\sum{\Delta f_1\otimes \Delta f_2\otimes
        \left(\hat{\sigma}^{f_1f_2\to f_3X}\hat{a}_{LL}^{f_1f_2\to f_3X}\right)\otimes D_{f_3}^h}}
     {\sum{f_1\otimes f_2\otimes\hat{\sigma}^{f_1f_2\to f_3X}\otimes D_{f_3}^h}}.
\label{factorizationDef}
\end{equation}
The $\otimes$ symbol denotes convolution and the summations run over the partons $f_1$ and $f_2$ from the
initial protons, as well as the scattered parton $f_3$ that fragments to the observable $h$. For jet
measurements, no fragmentation functions are needed. All quantities in the right hand side of Eq.
\ref{factorizationDef} are well constrained, except for $\helicity$, in particular in the $x<0.05$
region. Furthermore, measurements of inclusive jets, dijets, and pions at STAR kinematics are dominantly
sensitive to quark-gluon and gluon-gluon scattering subprocesses
\cite{subprocessFraction,centralSubprocessFraction,forwardSubprocessFraction}, therefore recent
measurements of $\asym$ at STAR are sensitive to the gluon helicity. 

This presentation focuses on measurements of $\asym$ at the STAR experiment at RHIC. Recent results from
three regions of pseudorapidity $\eta$ are presented: central rapidity with $|\eta|<1$, intermediate
rapidity with $0.8<\eta<2$, and forward rapidity with $2.65<\eta<3.9$. The data are from longitudinally
polarized proton-proton scattering at center-of-mass energies of $\sqrt{s}=200$ and $510$ GeV. These
measurements, in particular the forward rapidity measurements, are sensitive to gluon $x$ down to
${\sim}10^{-3}$. The next three sections summarize the results for these $\eta$ regions, beginning with
the central rapidity region.

\section{Central Rapidity Observables}
\begin{figure}[pt]
\centerline{\includegraphics[width=0.7\textwidth]{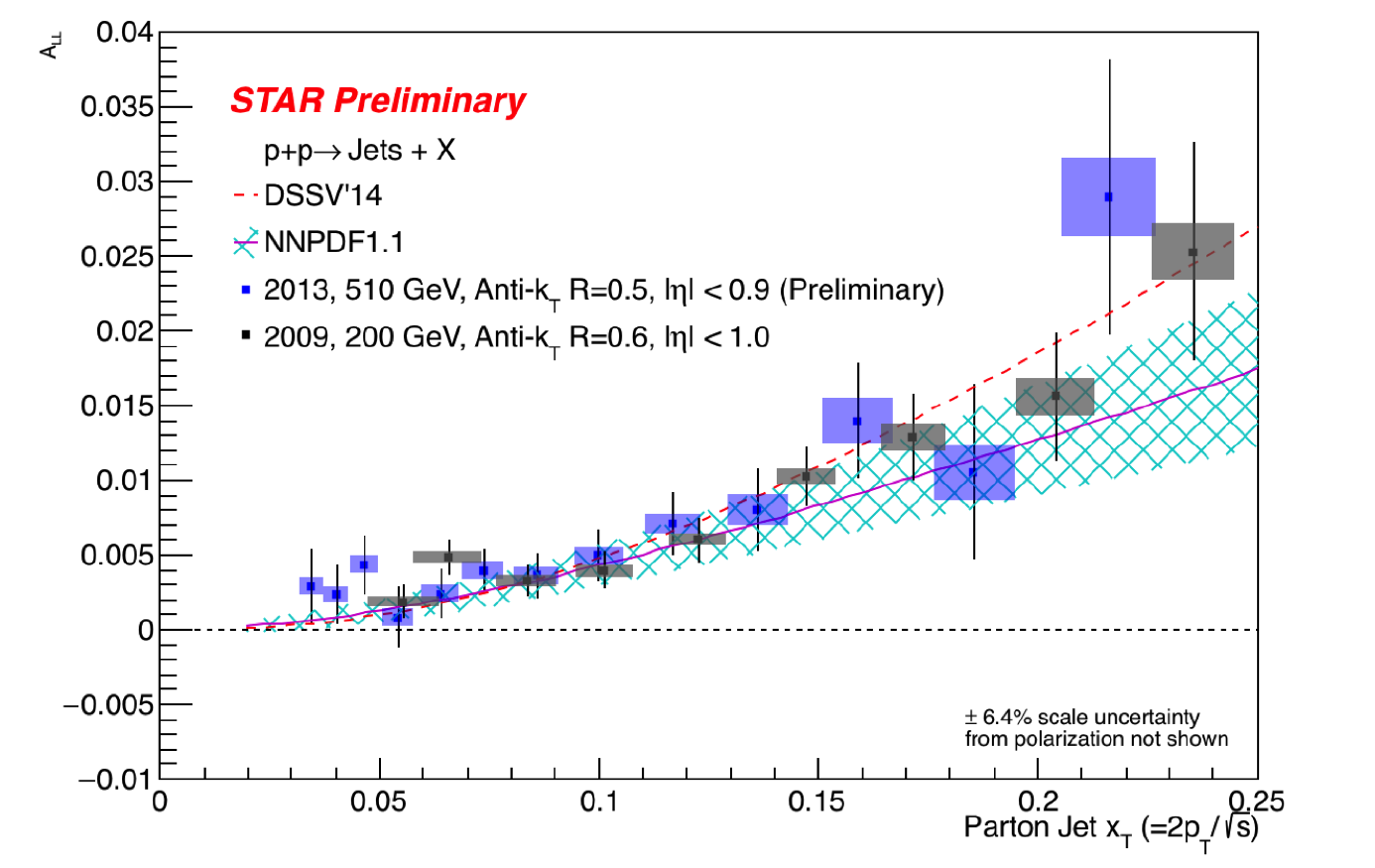}}
\caption{Comparison of central inclusive jet $\asym$ at $\sqrt{s}=200$ GeV from 2009 STAR data (black
points) \cite{centralJets200} to a preliminary measurement at $\sqrt{s}=510$ GeV from 2013 (blue points)
\cite{centralJets510}.
The vertical lines represent statistical uncertainties and the shaded boxes represent systematic
uncertainties. The DSSV14 \cite{dssv14} and NNPDFpol1.1 \cite{nnpdf1} global fit curves are shown as
well. From \cite{amilkarProceedings}.}
\label{centralJetsFig}
\end{figure}

Measurements of inclusive jets and dijets at central pseudorapidity, with $|\eta|<1$, are sensitive to
$x>0.05$ at $\sqrt{s}=200$ GeV and $x>0.02$ at $\sqrt{s}=510$ GeV. The inclusive jet analysis at
$\sqrt{s}=200$ GeV from 2009 STAR data \cite{centralJets200} yielded a positive $\asym$, systematically
above the DSSV08 global analysis fit \cite{dssv08}, which demonstrates that $\helicity$ for $x>0.05$ is
positive. Inclusive jet $\asym$ measurements at $\sqrt{s}=510$ GeV \cite{centralJets510} from 2012 and
2013 STAR data show agreement with the 200 GeV data, but with a sensitivity to somewhat lower $x$, down
to 0.02. This agreement is illustrated in figure \ref{centralJetsFig}, which shows a comparison of the
2009 $\sqrt{s}=200$ GeV and 2013 $\sqrt{s}=510$ GeV inclusive jet $\asym$ measurements.

\begin{figure}[pt]
\centerline{\includegraphics[width=\textwidth]{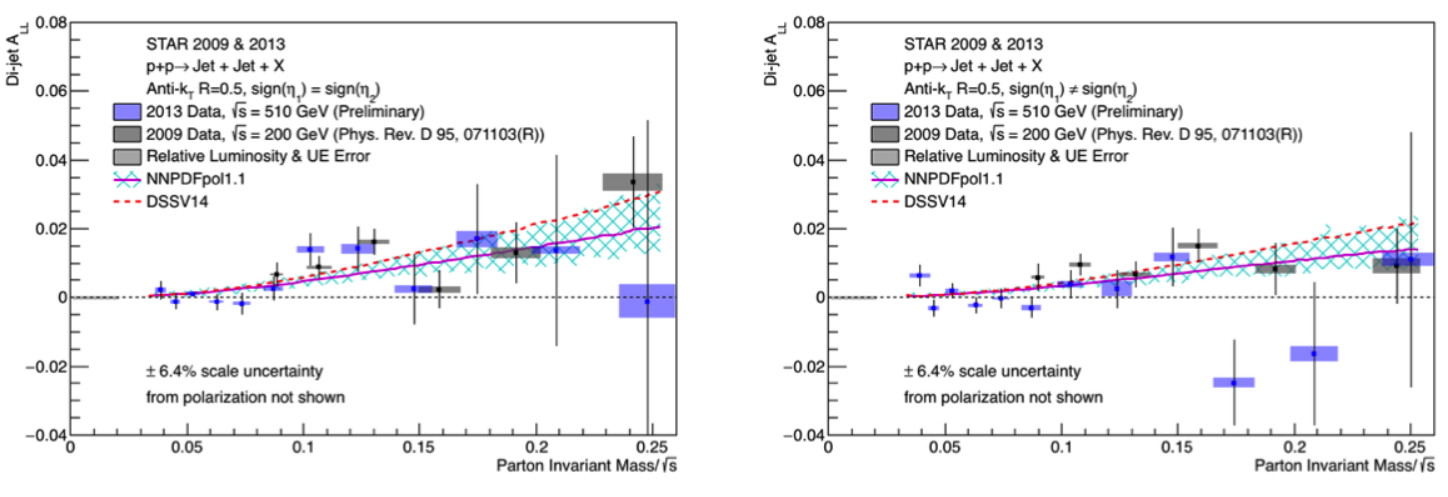}}
\caption{Comparison of central inclusive dijet $\asym$ at $\sqrt{s}=200$ GeV from 2009 STAR data (black
points) \cite{centralDijets200} to a preliminary measurement at $\sqrt{s}=510$ GeV from 2013 (blue
points).  The two panels are for the different dijet topologies: the left is for
$\mbox{sign}(\eta_1)=\mbox{sign}(\eta_2)$ and the right is for
$\mbox{sign}(\eta_1)\neq\mbox{sign}(\eta_2)$.  The vertical lines represent statistical uncertainties and
the shaded boxes represent systematic uncertainties. The DSSV14 \cite{dssv14} and NNPDFpol1.1
\cite{nnpdf1} global fit curves are shown as well. From \cite{amilkarProceedings}.}
\label{centralDijetsFig}
\end{figure}

Inclusive dijet measurements probe a much narrower region of $x$, because the partonic $x$ values are
calculable from the dijet kinematics. A measurement of dijet $\asym$ from $\sqrt{s}=200$ GeV 2009 data
\cite{centralDijets200} and a preliminary measurement from $\sqrt{s}=510$ GeV 2012-2013 data each show
positive asymmetries. These measurements agree with models that predict $\Delta G\sim0.2$, for two
different dijet topologies: the jets 1 and 2 both produced on the same side of the central transverse
plane ($\mbox{sign}(\eta_1)=\mbox{sign}(\eta_2)$), and each jet produced on opposite sides
($\mbox{sign}(\eta_1)\neq\mbox{sign}(\eta_2)$). Figure \ref{centralDijetsFig} shows a comparison of the
2009 $\sqrt{s}=200$ GeV measurement to that from 2013 $\sqrt{s}=510$ GeV data, for each of these 
jet topologies.

\section{Intermediate Rapidity Observables}
Moving forward in $\eta$ to the intermediate region, $0.8<\eta<2$, the sensitivity to gluon $x$ is pushed
down to $0.01$ for $\sqrt{s}=200$ GeV; analyses of observables in $\sqrt{s}=510$ GeV collisions will push
down to $x\approx 0.004$, but are still underway. The measurement of $\asym$ in inclusive dijets at $200$
GeV \cite{intermediateDijets} from 2009 data provides tighter constraints on the size and especially the
shape of $\helicity$. This is a newly published result and was presented at this conference; for further
details, see the presentation and proceedings by T. Lin \cite{tingProceedings}. 

Three dijet topologies were assessed: (1) one jet at intermediate
rapidity and the second jet at central rapidity, on the opposite side of the central plane as the first jet,
(2) one jet at intermediate rapidity and the second central jet on the same side of the central
plane, and (3) both jets at intermediate rapidity. Case (3) probes the lowest $x$ of the three topologies
and has a large $\asym$ systematically above the DSSV14 \cite{dssv14} and NNPDFpol1.1 \cite{nnpdf1}
theoretical predictions, whereas cases (1) and (2) show rather good agreement with theory. The analysis
of $510$ GeV dijets from 2012 and 2013 data, which will probe down to $x\approx 0.004$, is currently underway.

A somewhat older, but related result is the intermediate rapidity $\pi^0$ $\asym$ measurement from
$\sqrt{s}=200$ GeV data from 2006 \cite{intermediatePions}. This result agrees with the available models,
but is not significantly capable of distinguishing between them. Analysis of $\sqrt{s}=510$ GeV $\pi^0$
data is also underway.

\section{Forward Rapidity Observables}
Forward observables are sensitive to gluons of even lower $x$, since the dominant subprocess is a soft
low-$x$ gluon scattering off a hard mid-to-high-$x$ quark. The forward $\pi^0$ $\asym$ in the
pseudorapidity range $2.65<\eta<3.9$ probes gluon $x$ down to $10^{-3}$ \cite{forwardPions}. This
result has also been recently published, and some more details are provided here.

\begin{figure}[tp]
\centerline{\includegraphics[width=0.55\textwidth]{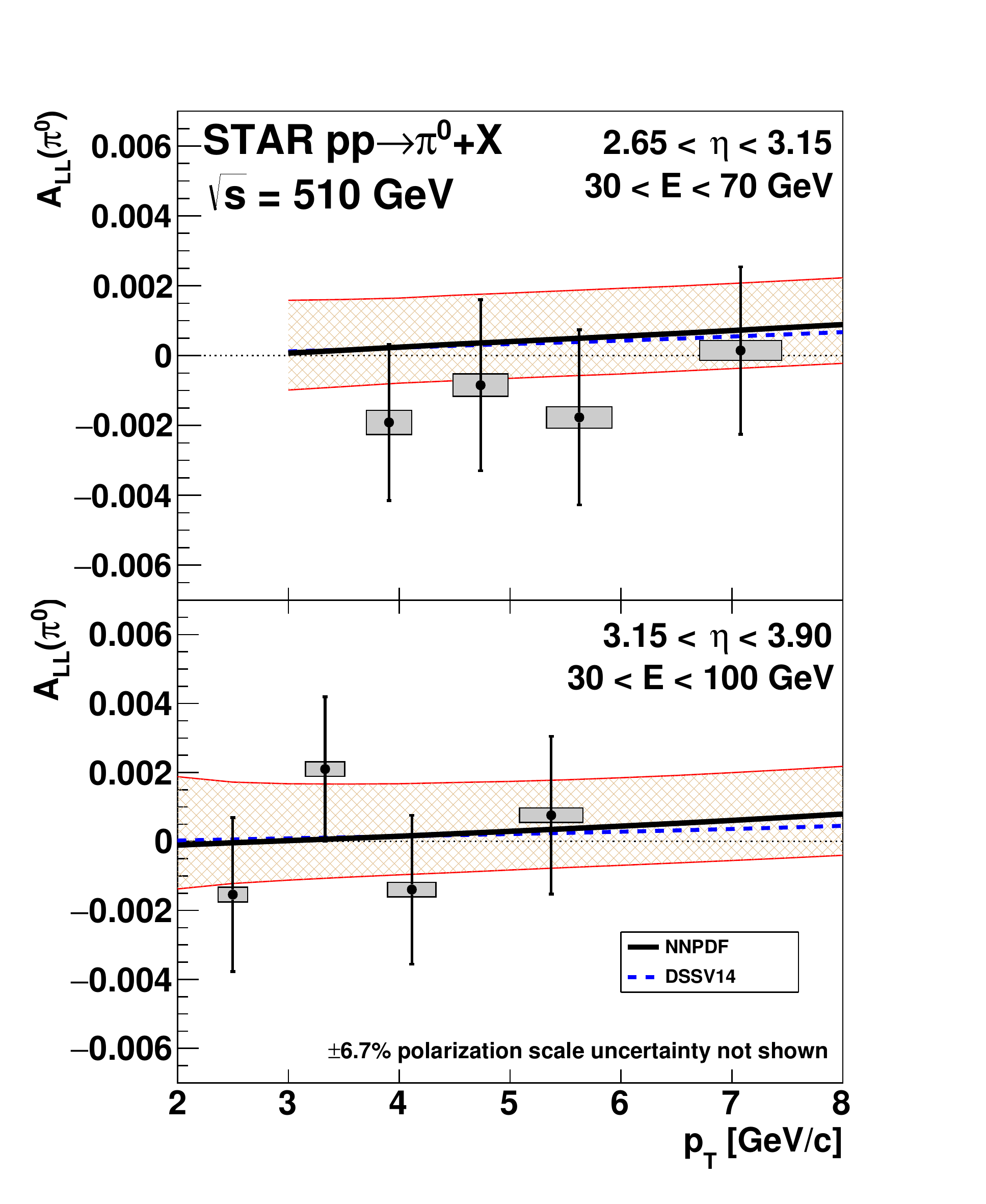}}
\caption{Forward $\pi^0$ $\asym$ for inner $\eta$ region (top) and outer $\eta$ region (bottom). See text
for details. From \cite{forwardPions}.}
\label{fig4}
\end{figure}

Figure \ref{fig4} shows the forward $\pi^0$ $\asym$ measurement, along with extrapolations of $\asym$
predictions to these forward kinematics, assuming the central $\helicity$ values from the recent DSSV14
\cite{dssv14} and NNPDFpol1.1 \cite{nnpdf1} global analyses. The top panel is for $\pi^0$ production in
the pseudorapidity range $2.65<\eta<3.15$ and the bottom panel is for the range $3.15<\eta<3.9$. The
vertical error bars on the data represent the statistical uncertainty, the horizontal extent of the
shaded gray boxes represents the systematic uncertainty on $p_T$, and the vertical extent represents the
systematic uncertainty on $\asym$; an additional 6.7\% scale uncertainty from the polarization is not
shown. 

The dashed blue curve represents the DSSV14 extrapolation, and the solid black curve along with the
hatched uncertainty band represents the NNPDFpol1.1 extrapolation. 
The uncertainty band was determined by calculating $\asym$ at a given $x$ for each of the 100 replica
PDFs, and taking the standard deviation of this distribution.
The data points agree with the extrapolated theory curves, within statistical and systematic
uncertainties. Nonetheless, they will help constrain the size of $\helicity$ down to $x{\sim}10^{-3}$ in
the upcoming global analysis. 

Looking ahead to the future, upgrade plans for a forward calorimeter and tracking system at STAR are
underway \cite{forwardUpgrade}; see also the presentation and proceedings by K. Barish
\cite{kenProceedings}.  The capability of measuring forward jets and dijets will also probe $x$ down to
${\sim}10^{-3}$, but will more precisely probe the shape of $\helicity$ in this important region.
Forward tracking detectors will be installed inside the STAR Time Projection Chamber for tracking forward
charged hadrons. A new electromagnetic calorimeter as well as a new hadronic calorimeter behind it will
take the place of the recently decommissioned Forward Meson Spectrometer, and when combined
with the new forward tracking system, will open the doors to many more detailed forward measurements.

\begin{table}[tp]
\resizebox{\textwidth}{!}{%
\begin{tabular}{?c|c?c|c?c|c?c|c?}
\hlineB{4}
\multirow{2}{*}{\textbf{$\sqrt{s}$ GeV}} & \multirow{2}{*}{\textbf{Year}}  &
\multicolumn{2}{c?}{\textbf{Central $\eta$}} &
\multicolumn{2}{c?}{\textbf{Intermediate $\eta$}} & \multicolumn{2}{c?}{\textbf{Forward $\eta$}} \\ \cline{3-8} 
                               &                        & \textbf{Jets}             & \textbf{Dijets}
                               & \textbf{Dijets}
                               & \textbf{Pions}              & \textbf{Pions}       & \textbf{Dijets}
                               \\ \hlineB{4}
\multirow{6}{*}{200}           & \multirow{2}{*}{2006}  & Published        &                  &
& Published          &             & \multirow{6}{*}{n/a}  \\ \cline{3-3}\cline{6-6}
                               &                        & $x>0.05$         &                  &
                               & $x>0.01$           &             &                       \\ \cline{2-7}
           & \multirow{2}{*}{2009}  & Published        & Published        & Published           &                    &             &   \\ \cline{3-5}
                               &                        & $x>0.05$         & $x>0.05$         & $x>0.01$
                               &                    &             &                       \\ \cline{2-7}
           & \multirow{2}{*}{2015}  & Underway         & Underway         &                     &
           & Underway    &   \\ \cline{3-4}\cline{7-7}
                               &                        & $x>0.05$         & $x>0.05$         &
                               &                    & $x>0.0025$  &                       \\ \hline
\multirow{4}{*}{510}           & \multirow{2}{*}{2012}  & Preliminary      & Preliminary      & Underway            & Underway           & Published   & \multirow{4}{*}{n/a}  \\ \cline{3-7}
                               &                        & $x>0.02$         & $x>0.02$         & $x>0.004$
                               & $x>0.004$          & $x>0.001$   &                       \\ \cline{2-7}
           & \multirow{2}{*}{2013}  & Preliminary      & Preliminary      & Underway            & Underway           & Published   &   \\ \cline{3-7}
                               &                        & $x>0.02$         & $x>0.02$         & $x>0.004$
                               & $x>0.004$          & $x>0.001$   &                       \\ \hlineB{4}
\multirow{2}{*}{200 and 510}   & \multirow{2}{*}{2021+} &                  &                  &
&                    &             & Future                \\ \cline{8-8} 
                               &                        &                  &                  &
                               &                    &             & $x>0.001$             \\ \hlineB{4}
\end{tabular}
}%
\caption{Summary of STAR $\asym$ analyses, sensitive to gluon helicity, along with $x$
sensitivity for each analysis.}
\label{summaryTable}
\end{table}

The next global analysis will likely include many of these newly published results. The
forward $\pi^0$s and intermediate dijets will help constrain $\helicity$ in the important low-$x$ region,
where the gluons are vastly dominant in the overall parton densities. The central jets and dijets will
help pin down the size and especially shape for $x>0.02$, and altogether a more accurate determination of
the gluon helicity contribution to the proton is expected in the near future.

\section{Summary}
Table \ref{summaryTable} summarizes the recent STAR $\asym$ measurement statuses and gluon $x$
sensitivities. The $\sqrt{s}=200$ GeV results are shown in the top half, followed by the 510 GeV results.
The columns are organized by pseudorapidity region, with observables within each region. For each
analysis, the progression of analysis status is from ``underway,'' meaning the analysis is being actively
worked on but no results have been released by STAR, to ``preliminary,'' meaning a first look at the
measurement has been released, but more work needs to be done before finalizing it, and finally to
``published,'' meaning the result has been finalized and published. The last row is for the future spin
program in 2021+ with the new forward upgrade, giving access to forward dijets.

\bibliographystyle{ws-ijmpcs}
\bibliography{sources}

\end{document}